\documentclass[aps,amssymb,amsmath, twocolumn, pra, superscriptaddress]{revtex4}
\usepackage{amsmath,amsfonts,amssymb,amstext,amscd,amsthm,dsfont}
\usepackage{physics}
\usepackage{color}
\usepackage{soul,xcolor}
\usepackage{gensymb}
\usepackage[normalem]{ulem}

\newcommand{\bla}{\color{black}} 

\usepackage[colorlinks, citecolor=blue]{hyperref}

\usepackage[sc,osf]{mathpazo}

\usepackage{graphicx}
\usepackage{bbold}
\usepackage{braket}
\usepackage{placeins}

\begin{document}
	
	\title{Enhanced non-Markovian behavior in quantum walks with Markovian disorder}
	\author{N. Pradeep Kumar}
	\affiliation{The Institute of Mathematical Sciences, C. I. T campus, Taramani, Chennai, 600113, India}
	
	\author{Subhashish  Banerjee}
	\affiliation{Indian Institute of Technology, Jodhpur, 342037, India}

	\author{C. M. Chandrashekar}%
	\email{chandru@imsc.res.in}
	\affiliation{The Institute of Mathematical Sciences, C. I. T campus, Taramani, Chennai, 600113, India}
	\affiliation{Homi Bhabha National Institute, Training School Complex, Anushakti Nagar, Mumbai 400094, India}
	
	\begin{abstract} 
		Non-Markovian quantum effects are typically observed in systems interacting with structured reservoirs. Discrete-time quantum walks are prime example of such systems in which, quantum memory arises due to the controlled interaction between the coin and position degrees of freedom. Here we show that the information backflow that quantifies memory effects can be enhanced when the particle is subjected to uncorrelated static or dynamic disorder. The presence of disorder in the system leads to localization effects in 1-dimensional quantum walks. We shown that it is possible to infer about the nature of localization in position space by monitoring the information backflow in the reduced system. Further, we study other useful properties of quantum walk such as entanglement, interference and its connection to quantum non-Markovianity.  
	\end{abstract}
	\maketitle 
	
	\section{Introduction}
	Quantum walks  describe the coherent evolution of a quantum particle coupled to an external environment which in simple form is can be a position space. It is formally the quantum analogue of classical random walk \cite{Ria1958,Fey86,Mey96,ADZ93,Par88}. The ability of quantum walks to exploit certain non-trivial quantum effects such as interference and entanglement has led to a plethora of application in quantum information processing tasks \cite{Kem03,Ven12}. The most important among these is the capability to perform universal quantum computation \cite{Chi09,LCE10} and exponential speed up \cite{CDF03,CG04} to some quantum algorithms \cite{Amb07,MSS07,BS06,FGG08}. Another important application of quantum walks is quantum simulations. The inherent controllable nature of these systems allows us to gain a wealth of knowledge by simulating quantum phenomenon such as photosynthesis \cite{GCE07}, relativistic quantum effects \cite{Mey96,Fwa06,CBS10,Cha13,MBD13,MBD14,AFF16,Per16,MMC17,MP16}, localization \cite{Joy12,Cha12,OK11,KRB10,CB15}, among many others. In addition to this, the rich dynamical structure of quantum walks serves as a test bench for studying a wide class of open quantum system dynamics \cite{BPbook,CRB07,BSC08}.
	
	With progress in the development of quantum coherent devices and the rapid advancement in controlling these systems effectively, we are now witnessing both proof of principle experiments of quantum walks and performance of useful quantum simulations in different physical systems such as cold atoms \cite{KFC09}, ion traps \cite{SMS09,ZKG10} and in circuit-QED architectures \cite{FRH17}. In view of these recent developments in engineered quantum systems it is imperative to carefully study the interplay of various quantum features and its effects on quantum dynamics. In quantum walks which evolves in position space, the reduced dynamics of the particle obtained after tracing out the position degrees of freedom has been shown to exhibit certain features of quantum non-Markovianity such as information backflow \cite{BLPV16, HFR14}.  Recently, in \cite{HFR14,PSS17,TXX17} the effect of decoherence caused by the interaction of the particle with an external environment or due to the presence of broken links in position lattice has been shown to reduce the effects of non-Markovianity. Furthermore, in \cite{MSQ17} classical non-Markovian random process such as the Elephant random walk has been generalized to the quantum setting.
	
	In this work we study the intricate connections between quantum interference, entanglement and dynamical properties like non-Markovian quantum effects arising in the time evolution of quantum walks \cite{PSS17, TXX17, HFR14, MSQ17}. While quantum interference is understood to be the most fundamental resource that powers quantum walks, the complexity in tracking the changes and measuring it unambiguously has prevented a direct handle for studying it in quantum systems. However, recent progress \cite{SCMC17} has been made to estimate interference in quantum walks, which will be used here to make a comparative study of other resources such as entanglement and memory effects in quantum walk evolution due to non-Markovian dynamics. 
	
	The major roadblock in the large scale implementation of quantum walks is the inevitable presence of decoherence caused by the ubiquitous coupling between the system and the environment. In addition to this, the presence of unavoidable systematic errors that arises due to imperfect control operations generic to many quantum information processing systems \cite{MEH18,EHH18} also  impends the scalability of quantum walks. It is hence indispensable to understand the impact of these non-idealities on the dynamical properties of quantum walks. In view of that, we study the interplay of entanglement and non-Markovian memory effects  in the presence of static and dynamic disorder. This in turn leads to either  Anderson localization (spatial) or weak localization (temporal ) of the walker in the position space. An interesting connection between localization and quantum non-markovianity has been studied in the context of continuous-time quantum walks \cite{BBP16}. The results show that Anderson localization appears only when the evolution is subjected to non-Markovian regime of the noise. Alternatively, it was identified that Anderson localization induces non-Markovian features in dynamics \cite{LLC17}. In this work we show that, by introducing uncorrelated time and position dependent disorder in discrete-time quantum walk evolution, the memory effects in the system can be enhanced.  This further allows us to probe the nature of localization by observing only the reduced dynamics of the particle. \\
		
	\noindent 
	\section{Quantum walks} 
	Quantum walk on 1-D lattice is defined on the Hilbert spaces $\mathcal{H}_c$ spanned by the coin basis states which serve as the internal degrees of freedom $\ket{\uparrow}$ and $\ket{\downarrow}$, and $\mathcal{H}_p$ represents the external degrees of freedom, such as the position state which is spanned by the basis states $\ket{x}$, where $x \in \mathbb{Z}$. The state of the complete quantum system will be represented on the tensor product space $\mathcal{H}_c \otimes \mathcal{H}_p$ and the initial state will be in the form,
	\begin{equation}
	\ket{\Psi(0)}  = \text{cos}(\delta) \ket{\uparrow} + e^{-i\eta} \text{sin}(\delta) \ket{\downarrow} \otimes \ket{x=0}.
	\end{equation}
	Here, $\delta, \eta$ specifies an arbitrary initial state of the two level particle (coin) and the the position state $\ket{i=0}$ denotes the origin of the walker. The time evolution for the quantum walk is generated by the unitary transformations, coin operation followed by the shift operator. The action of the coin operator   
	\begin{equation}
	\hat{C}_\theta=
	\begin{bmatrix}
	~~\cos\theta & -i\sin\theta \\ \label{coin}
	-i\sin\theta & ~~\cos\theta
	\end{bmatrix},
	\end{equation}
	is only on the coin subspace and drives each internal state of the particle to superposition states. The coin operator specified above can be represented by a combination of Pauli matrices $C_{n, \theta} = e^{-i\theta \hat{\bf n} \vec{\sigma}}$, where $\hat{\bf n}$ is the unit vector which defines the direction. Here we have chosen $\hat{\bf n} = \hat{n_x}$ and hence $C_{x, \theta} = e^{-i\theta\sigma_x}$, as defined in Eq.\,(\ref{coin}), rotates the internal state of the particle in the $y-z$ plane of the Bloch sphere by the specified angle $\theta$. Depending on the internal state of the particle, the shift operator,
	\begin{equation}
	\hat{S}=\ket{\downarrow}\bra{\downarrow}\otimes\sum_{x\in\mathbb{Z}}\ket{x-1}\bra{x}+\ket{\uparrow}\bra{\uparrow} \otimes 	\sum_{x\in\mathbb{Z}}\ket{x+1}\bra{x}, 
	\label{shift}
	\end{equation}
	shifts the particle in superposition of position space. The overall unitary operator applied at every time step is $\hat{U}_\theta = \hat{S}(\hat{C}_\theta \otimes I)$ and the state at any time $t$ will be given by, 
	\begin{equation}
	\ket{\Psi(t)} = \hat{U}^t_\theta \Psi(0) = \hat{U}^t_\theta [\ \ket{\psi (\delta, \eta)} \otimes \ket{x=0}].
	\label{qeve}
	\end{equation}
	\bla
	\begin{figure}
		\includegraphics[scale=0.90]{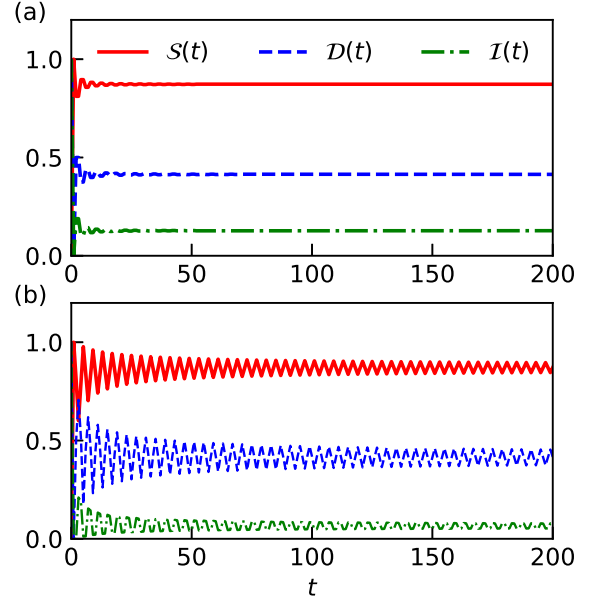}
		\caption{(Color online)Plot of Entanglement $S(t)$, trace distance $D(t)$ and interference $I(t)$ as a function of time for different initial states of the particle (a) $\ket{\psi(\frac{\pi}{4},0)}$ is chosen as an initial state for both $S(t)$ and $I(t)$, for $D(t)$, the initial pair of states are $\ket{\psi(\pm\frac{\pi}{4},0)}$. (b) The initial state here is $\ket{\psi(\frac{\pi}{4},\frac{\pi}{2})}$ for entanglement and interference, while $\ket{\psi(\pm\frac{\pi}{4},\frac{\pi}{2})}$ is used for computing $D(t)$. The coin parameter is set to $\theta = \frac{\pi}{4}$ for both the plots. We observe the oscillatory nature of all the three curves, typical for non-Markovian evolutions. The trace distance between initial states are bounded between entanglement and interference.}
		\label{fig1}
	\end{figure}	
	\bla

	\section{Results}
	\noindent
	\textit{Entanglement, Interference and non-Markovianity -}
	Our primary interest is to study the reduced dynamics of the two level particle, in the coin space, that is coupled to the position degrees of freedom. To understand the intricate features in the quantum dynamics in the coin space we have computed quantities such as entanglement, interference and non-Markovianity using suitable measures. The entanglement of the particle with the position space is measured using Von-Neumann entropy, 
	\begin{equation}
	S^{\rho_c}(t) = \text{tr}[\rho_c(t) \ \text{log}_2\{\rho_c(t)\}]
	\end{equation}
	where $\rho_c(t) = tr_p[\rho(t)]$ is the reduced state of the particle. 
	
			\begin{figure}
				\centering
				\includegraphics[scale=.90]{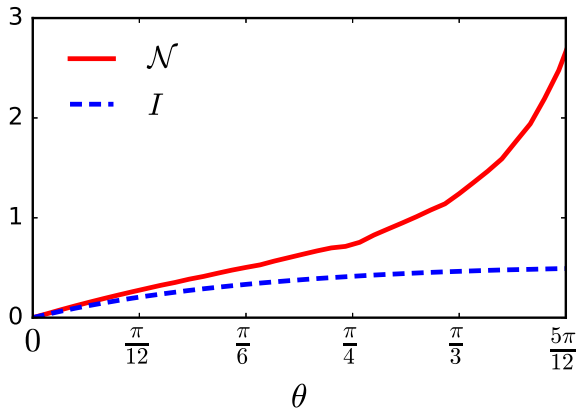}
				\caption{(Color online) BLP measure and interference as a function of the coin parameter $\theta$ computed after 200 steps of quantum walk for the initial coin state $\ket{\psi(\frac{\pi}{4},0)}$. Both non-Markovianity and interference increases monotonically with increase in $\theta$ sampled in the interval $\big[0, \frac{5\pi}{12}\big]$.}
				\label{fig2}
			\end{figure}\bla

	To compute the amount of interference present in the coin space, we can use the coherence measure, which is the absolute sum of the off-diagonal elements of the reduced state of the particle \cite{BCP14,SCMC17}, 
	\begin{equation}
	I^{\rho_c}(t) = \sum_{i\neq j} |\rho^{ij}_c(t)|.
	\end{equation}
	In Fig.\,\ref{fig1}(a) \& (b) we have plotted the entanglement ($S(t)$) and interference ($I(t)$) in coin space as function of time for different initial coin states. We notice that for a given initial state, the value of entanglement is significantly higher than the value of interference but the patterns are identical. However, irrespective of the initial state, the mean value of entanglement and interference is same in both Fi g.\,\ref{fig1}(a) \& (b). In Fig.\,\ref{fig1}(b) the oscillatory nature of the curves is prominently visible, indicating the strong presence of memory effects due to the information backflow \cite{BLP10} between the coin and position degrees of freedom. In Fig.\,\ref{fig1}(a) oscillation vanishes very quickly with time. In order to witness the information backflow we resort to the trace distance between reduced density matrix evolved with different initial states.  Trace distance between any two density matrices is defined as, 
	\begin{equation}
	D(\rho_1(t), \rho_2(t)) = \frac{1}{2} \mbox{Tr}|\rho_1(t) - \rho_2(t)|,
	\end{equation}
	where $\rho_1 (t)$ and $\rho_2(t)$ are the reduced density matrix after time $t$ starting from two different initial state of the particle. In Fig.\,\ref{fig1}, trace distance for two difference combination of initial states is shown and we note that trace distance essential captures the same behavior as that of the entanglement and interference with different mean value. However, trace distance plays an important role in quantifying the non-Markovianity in the dynamics.
	
	The non-Markovianity of the reduced dynamics of the particle can be quantified using the Breuer, Laine, Pillo (BLP) measure \cite{BLP10}. The BLP measure is defined as follows, 
	\begin{equation}
	\mathcal{N} = \int_{\sigma > 0} \sigma(t, \rho^{1,2}(0)) \ dt,
	\label{blpm}
	\end{equation}
	where $\sigma(t, \rho^{1,2}(0)) = \frac{d}{dt}D(\rho^1(t), \rho^2(t))$ is the derivative of the trace distance between the reduced density matrix $\rho^1(t), \rho^2(t)$ obtained using two initial states $\ket{\psi(\pm\frac{\pi}{4}, \frac{\pi}{2})}$.
	
	BLP measure captures the information flow between the position and the particle degrees of freedom, using the notion of distinguishably of quantum states. We should remark that that the BLP measure is originally defined as the maximization over initial states in order to obtain the maximum possible non-Markovianity for the given dynamical map. Since we are not interested in this we have dropped the maximization term in Eq.\,(\ref{blpm}).  However, it is useful to mention that in the quantum walk dynamics, Eq.\,(\ref{blpm}) is maximized for the initial states $\ket{\psi(\pm\frac{\pi}{4}, \frac{\pi}{2})}$ as shown in \cite{HFR14}.
	
	The non-Markovian behavior characterized by trace distance correlates well with interference. The direct relationship between interference and non-Markovianity can be observed in Fig.\,\ref{fig2}. We notice that the BLP measure ($\mathcal{N}$) monotonically increases with $\theta$ which in turn controls the amount of interference. \\
		
	\noindent 
	\textit{Enhancement of non-Markovianity in the presence of disorder-}
	In the simple scenario where the quantum coin operator is fixed, the time evolution is homogeneous and the evolution follows Eq.\,(\ref{qeve}). The quantum walk can be made inhomogeneous by introducing disorder in the system which breaks the time translation symmetry of the evolution. This leads to the localization effects in the position basis states which inhibits the spread of the walker. We classify disorder into two types, dynamic and static, depending on whether the coin operator changes randomly with time step or position respectively. \\
	
	\begin{figure}
		\includegraphics[scale=0.90]{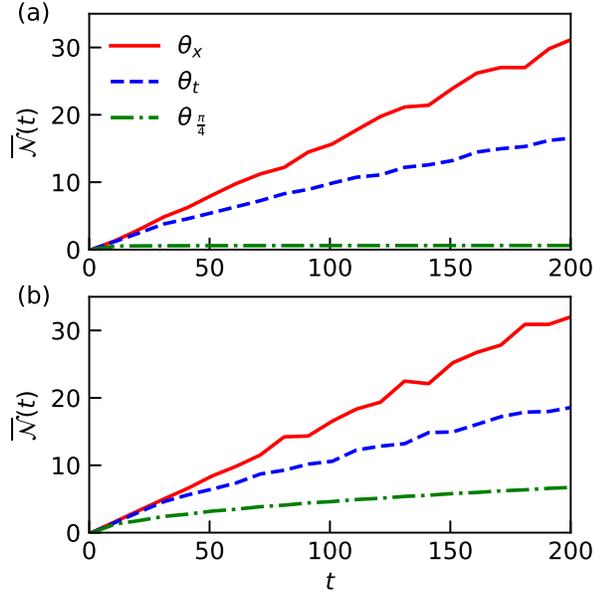}
		\caption{(Color online) The amount of non-Markovianity quantified using BLP measure after every step of the quantum walk for different coin parameter $\theta$ for two different initial state of the particle. (a) $\ket{\psi(\pm\frac{\pi}{4}, 0)}$ (b) $\ket{\psi(\pm\frac{\pi}{4}, \frac{\pi}{2})}$. The computations are averaged over $10^3$ simulations for 100 percent disorder strength. We observe the enhancement of non-Markovianity for both spatial and temporal case, irrespective of the initial states.}
		\label{fig3}
	\end{figure}
	We consider two independent and identically distributed random variables $\tilde{\theta}(t)$ and $\tilde{\theta}(x)$ representing the functional dependence on the time or position, respectively. The underlaying probability distribution is considered to be uniform. The random process generated by $\tilde{\theta}(m)$ where $m = t, x$ is a white noise process which has the delta correlation function of the form, 
	\begin{equation}
	\langle \tilde{\theta}(m)  \ \tilde{\theta}(m') \rangle = \delta_{mm'}.
	\label{cf}
	\end{equation} 
	The above correlation function holds in general for any classical Markovian process. The effect of disorder essentially causes errors in the qubit  (state of the particle) rotation due to random coin operations $e^{-i\tilde{\theta}(m)\sigma_x}$. This type of imperfect rotations are often considered as ``classical noise" while modelling quantum gate errors \cite{MEH18,EHH18} as opposed to quantum noise arising due to genuine system bath interaction which affects the internal state of the particle  \cite{BPbook}. 
	
	The temporally disorder quantum walks evolve according to the following equation,
	\begin{equation}
	\ket{\Psi(t)} = \left(\hat{U}_{\tilde{\theta}(t)}\right)^t \ket{\Psi(0)},
	\end{equation}
	where, $\hat{U}_{\tilde{\theta}(t)}$ is the effective unitary operator that encodes the time dependent random coin operation, sampled in the interval [0, $\frac{\pi}{2}$]. The corresponding trajectory is, 
	\begin{equation}
	\ket{\Psi(t)} = \hat{U}_{\theta(n)} \hat{U}_{\theta(n-1)}.........\hat{U}_{\theta(1)}\ket{\Psi(0)}.
	\end{equation}
	Here $n$ denotes the particular time instance of the quantum walk evolution. 
	
	Similarly, spatial disorder can be introduced in quantum walk by applying a position dependent coin operation which mimics the disorder in the position degrees of freedom \cite{Cha12},
	\begin{equation}
	\ket{\Psi(t)} = \left(\hat{U}_{\tilde{\theta}(x)}\right)^t \ket{\Psi(0)}.
	\end{equation}
	Here $\hat{C}_{\tilde{\theta}(x)} = \sum_{x} \hat{C}_{\theta (x)} \otimes \ket{x} \bra{x}$ is the position dependent coin operation that is encoded in the effective unitary operator $\hat{U}_{\tilde{\theta}(x)}.$  Both, temporal and spatial disorder are detrimental to certain applications of quantum walks like search algorithms. However, by tailoring the disorder 
	process it can play a significantly role in simulating the exotic dynamics of disordered materials.
	\begin{figure}
		\includegraphics[scale=0.90]{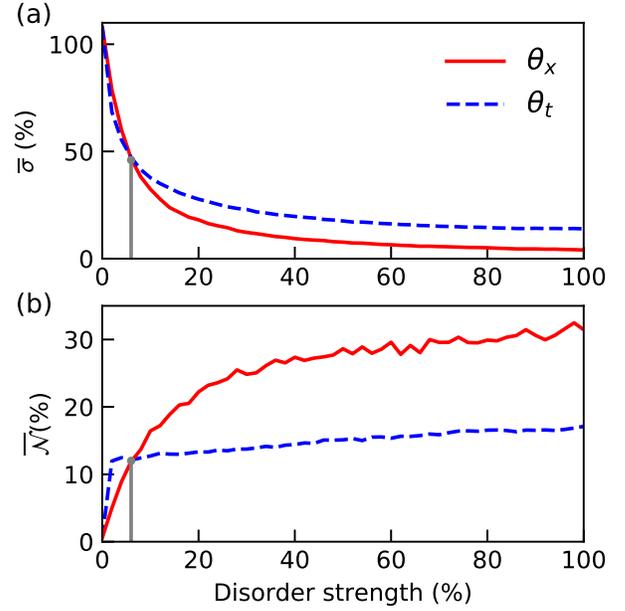}
		\caption{(Color online) Plot of standard deviation and BLP measure as a function of the strength of the disorder used to witness the enhancement of non-Markovian memory effects averaged over $10^3$ simulations. (a).Standard deviation of the position probability  distribution. (b).  BLP measure that quantifies non-Markovianity. Enhancement of non-Markovianity is to be noted for both spatial and temporal disorder. Identical threshold point for both the plots are denoted in Grey solid line.}
		\label{blpdis} \bla
	\end{figure}
	\begin{figure}
		\includegraphics[scale=0.85]{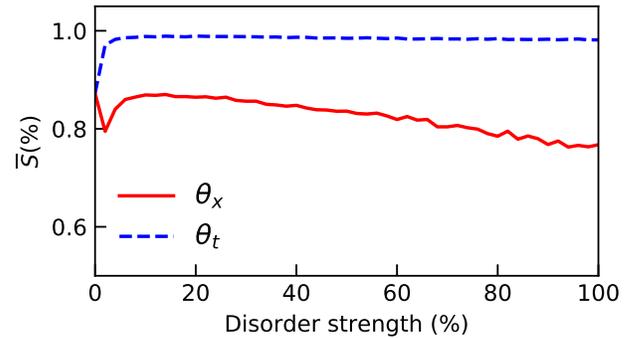}
		\caption{(Color online) Average Entanglement between the coin and position degrees of freedom plotted as a function of disorder strength evaluated after 200 steps and computed by averaging over $10^3$ simulations. The effects of temporal disorder enhances the entanglement while the spatial disorder tend to decreases it.}
		\label{entdis}
	\end{figure}

			
	The effect of both the temporal and the spatial disorder leads to localization of the walker in external degrees of freedom of the particle. While temporal disorder results is a weaker form of localization, spatial disorder leads to Anderson localization. The connection between non-Markovianity and disorder has been explored by studying a microscopic model of a two level atom coupled to a circular lattice \cite{LLC17}.  It has been explicitly shown that Anderson localization induces non-Markovianity, since the disorder effectively introduces strong coupling between the system and a few set of environmental modes. In the quantum walk scenario similar effects can be reproduced. We should note that the non-Markovian effects are generated between the internal and external degrees of freedom even when there is no disorder in the quantum walks evolution. This is shown in Fig.\,\ref{fig3} for different initial pair of states with the coin parameter set to $\theta = \frac{\pi}{4}$. We observe that,  when the quantum walk is made inhomogeneous by introducing disorder into the system, the memory effects in the evolution is enhanced from the homogeneous case. We explicitly show this by ensemble averaging the BLP measure $\overline{\mathcal{N}}$ over large simulations, both as a function of time and disorder strength. From Fig.\,\ref{fig3} we observe that, $\overline{\mathcal{N}}$ increases with the strength of disorder for both  temporal and spatial disorder. However, spatial disorder  tends to enhance non-Markovianity more in comparison to temporal disorder that almost saturates as the disorder strength increases. Another interesting feature that results from this study is the indication of the existence of a threshold value that differentiates spatial and temporal localization in terms of the disorder strength. 
	
	In order to verify our results we chose to compute the ensemble averaged standard deviation ($\overline{\sigma}$) in the position space, which is a more intuitive metric to study the effect of disorder in the system. We note that a similar threshold point can be found in Fig. \ref{blpdis} (a)  that differentiates temporal and spatial disorder. This is corroborated by the behavior of the ensemble averaged BLP measure depicted in Fig.\ref{blpdis} (b). This is interesting for two reasons; firstly the qualitative results in the microscopic model studied in \cite{LLC17} are reproduced in the quantum walks dynamics, and secondly the localization effect that appears in the position space in effectively captured by analyzing the dynamical behavior of the reduced system, that is, state of the particle alone. 	 
		
	It is interesting to note that both spatial and temporal disorder result in enhancement of non-Markovianity whereas enhancement of entanglement is seen only with temporal disorder (Fig.\,\ref{entdis}). This intriguing behavior needs further investigations which can result in better understanding of quantum correlations and non-Markovianity in disordered systems.   Our studies also show that a small amount of disorder is sufficient to bring about a significant rise in non-Markovianity of the dynamics. \bla
	
		\begin{figure*}[htpb]
			\includegraphics[scale=1.05]{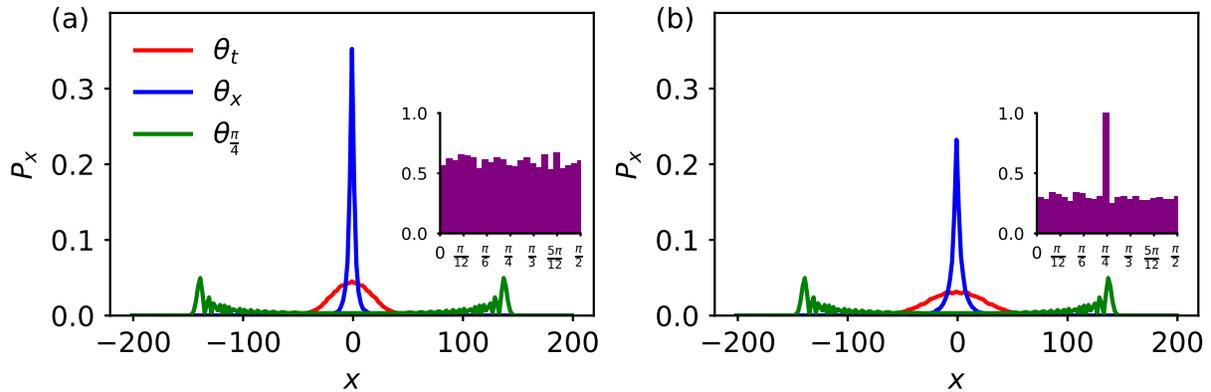}
			\caption{(Color online) The probability  distribution of the quantum walk with different types of random coin operations. (a). The strength of disorder is $100 \%$ for both temporal and spatial disorder, the inset plot shows the histogram of random coin operations sampled uniformly over the interval $[0, \frac{\pi}{2}]$. (b). The strength of disorder is $50\%$, leading to localization weaker than (a), the inset plots shows the histogram with a peak at $\theta = \frac{\pi}{4}$ chosen predominantly over the rest of the   values.}
			\label{pdis}
		\end{figure*}

	\section{Methods}
	In the numerical simulations, the quantum walk is evolved up to 200 steps, sufficient to support all the results.  The random sequence of coin operations for the disordered quantum walk is generated from a uniform probability distribution. It is important to note that the enhancement of non-Markovianity  is observed independent of the shape of the probability distribution, the only necessary condition is that for the random variable $\theta$ to obeys Eq. (\ref{cf}). The strength of disorder given in $\%$, is modelled as the average number of times the random coin operator appears in either,  total number of time steps for temporal disorder or number of position basis states for spatial disorder. The homogeneous quantum walk is recovered when the  disorder strength is zero and the maximally disordered quantum walk is obtained when the disorder strength is $100 \%$, that is when all the coin operators flips randomly. The effect of disorder strength on the position probability distribution is shown in Fig. \ref{pdis}.  The effect of localization has been measured using two different metrics.  (1)The discrete version of the non-Markovianity measure described in Eq. (\ref{blpdis}), which was introduced in \cite{BLP10} and also studied in the context of quantum walks \cite{LP16}, (2) is the standard deviation measure defined as $\sigma = \sqrt{\langle P_x^2\rangle - \langle P_x\rangle^2} \ $, where $P_x$ is the probability of finding the walker at position x. 
			
	\section{Discussion and conclusion}
	We have shown that the Markovian disorder in the form of white noise in quantum walks enhances non-Markovian behavior in the dynamics.  In particular, we have shown that the increase in non-Markovian behavior is noticeably higher for spatial disorder which results in Anderson localization when compared to temporal disorder which leads to weak localization. With this we have shown that the nature of localization in the position space can be analyzed from its non-Markovian behavior obtained by monitoring the information backflow  in reduced system. The non-Markovian nature of the quantum walk dynamics can be viewed as a potential resource, which can be exploited in various metrological applications to probe the dynamics of complex many body systems \cite{HM14}.  
	
	Another interesting observation that is a fall out of this work is the contrasting behavior of entanglement  and non-Markovianity in the presence of spatial disorder. While both spatial and temporal disorder can enhance information backflow,  enhancement of entanglement between the position and coin degrees of freedom is possible only with temporal disorder. These observations need further investigations to discern the usefulness of different types of disorder that can be engineered to enhance different properties of the system. \bla


\vskip 0.2in
	\noindent
	{\bf Acknowledgment:}\\
	\\
	\noindent
	CMC and NPK would like to thank Department of Science and Technology, Government of India for the Ramanujan Fellowship grant No.:SB/S2/RJN-192/2014. NPK acknowledges useful discussions with Nana Siddarth and Geroge Thomas.

\vskip 0.2in
	
\noindent  
{\bf Author Contributions:}\\

\noindent
CMC and SB conceptualized the study. NPK and CMC carried out the numerical analysis and prepared the figures. All the three authors were involved in interpreted the results and writing of the manuscript.

\vskip 0.2in
\noindent
{\bf Additional Information :}\\

\noindent
{\bf Competing financial interests:} The authors declare no competing interests.

\end{document}